\documentclass[aps,preprint,a4paper,showpacs,amsmath,amssymb,floatfix]{revtex4}
\usepackage{graphicx}% Include figure files
\usepackage{dcolumn}% Align table columns on decimal point
\usepackage{bm}
\newcommand{\Op}[1]{{{\mathrm{\hat{#1}}}}}

\begin{document}
%\draft
\title{The interplay between cycle geometry and performance of  {\em sudden \rm}  refrigerators}

\author{Tova Feldmann and Ronnie Kosloff }

\address{Institute of Chemistry, 
Hebrew University of Jerusalem, Jerusalem 91904, Israel}
%\ead{ronnie@fh.huji.ac.il}

\begin{abstract}
The relation between the geometry of refrigeration cycles and their performance is explored.
The model studied is based on a coupled spin system. 
Small cycle times termed {\em sudden} refrigerators, develop coherence and inner friction. We explore the interplay 
between coherence and energy of the working medium employing a family of sudden cycles with decreasing cycle times.
At the point of minimum ratio between energy and coherence the cycle changes geometry.
This region of cycle times is characterised by short circuit cycles where heat is dissipated both to the hot and cold baths.
We rationalise the cycle  change of geometry as a result of a half integer quantization which maximises coherence.
From this point on, increasing or decreasing the cycle time, eventually leads to refrigeration cycles. 
The transition point between refrigerators
and short circuit cycles is characterised by a transition from finite to singular dynamical temperature.
Extremely short cycle times reach a universal limit where all cycles types are equivalent.
\end{abstract}
\maketitle

\section{Introduction}
\label{sec:intro}

The study of quantum heat engines is motivated by the quest to understand the fundamental relation
between quantum mechanics and thermodynamics \cite{k281,alla11,popescu10,quan2007}. 
In addition the promise of quantum technologies 
requires a detailed understanding of quantum devises. 
Typically, quantum devices operate at ultracold conditions  
therefore miniature refrigeration is a prerequisite for successful application. A lingering question  in the field is:
what is quantum in the operation of quantum  refrigerators? Attempts to link the optimal
performance point to entanglement yielded inconclusive results \cite{brunner2014,correa2014}.

The present study is devoted to elucidating the quantum character of reciprocating quantum refrigerators. 
In particular the quantum character of the limit cycle of refrigerators operating far from equilibrium.
Analysis will allow to generalise the phenomena  to more realistic devices.
The prototype of quantum reciprocating cycle is the Otto cycle \cite{k116,k152,k176,k190,k243,k251,k258,thomas11,albayrak2013,huang2014}.
The cycle is composed of two isochores  where the external force is constant and two adiabats
where the Hamiltonian changed its energy scale induced by an external parameter variation.
For slow variation of the parameter, quantum adiabatic conditions can be achieved. Then
at all times the working medium is diagonal in the energy representation.
Fast variations generate energy excitation and reduced efficiency. For optimal efficiency and cooling power
it is sufficient that the working medium is diagonal in the energy representation 
at the beginning and at the end of the adiabatic stroke. 
These solutions are termed frictionless or shortcuts to adiabaticity \cite{del2013shortcuts,jarzynski2013generating,k251}.  For such cycles  
the whole operation  can be described by energy level populations hiding any 
quantum feature. Such refrigerators are described as stochastic \cite{schmiedl2008}.

All heat engine and refrigerator cycles are subject to a global constraint: the cycle must close on itself. 
As a result all observables of the working medium have the same periodic character. 
We will show that this topological constraint has a profound influence on the performance.
In the stochastic limit, the cycle imposes only a constraint on the internal energy. 
In the Otto cycle entropy changes of the working medium occur  only in the thermalization strocks 
therefore the entropy change on the hot and cold strokes have to be exactly balanced. 
These constraints position the cycles on the entropy energy plane \cite{k152}.

The present study is devoted to refrigerators which operate with large coherence. 
Coherence is generated for short cycle times, shorter than the characteristic timescale of the working medium.
We term these cycles as {\em sudden}. 
For these cycles the topological constraint involves both energy and coherence,
influencing both the geometry  and performance of the cycle.

\section{Quantum Otto Refrigerator Cycles}
\label{general}

The current  study is based on the quantum Otto cycle with  coupled spin system as a working medium.
The details of the model can be found in previous studies \cite{k176,k190,k201,k251,k258,k274}.
Only relevant features will be outlined in the present paper.
The Otto cycle studied is composed of the following four segments (Cf. Fig. \ref{fig:1}):

\begin{enumerate}

\item {Segment  $A \rightarrow B$ ( \em isomagnetic \rm or  \em isochore\rm), partial equilibration with the cold bath under constant Hamiltonian. The dynamics of the working medium is characterised by the  propagator ${\cal U}_c$.}

\item   {Segment ~$B \rightarrow C$ ( \em magnetization \rm or \em compression adiabat\rm),
the external field  changes expanding the gap between energy levels of the Hamiltonian.
The dynamics is characterised by the propagator ${\cal U}_{ch}$.}

\item   {Segment ~$C \rightarrow D$ (\em isomagnetic \rm or \em isochore\rm) partial 
equilibration with the hot bath described by the  propagator ${\cal U}_h $.}

\item   {Segment ~$D \rightarrow A$ (\em demagnetization \rm or 
        \em expansion adiabat \rm) reducing the energy gaps in the Hamiltonain, 
         characterised by the  propagator ${\cal U}_{hc}$.}

\end{enumerate}
The propagator of the four stroke cycle 
becomes ${\cal U}_{global}$,  which is the ordered product of the segment propagators:
\begin{equation}
{\cal U}_{global}~~=~~        {\cal U}_{hc} {\cal U}_h {\cal U}_{ch} {\cal U}_c      
\label{eq:globalprop}  
\end{equation} 
The propagators are linear operators defined on a vector space which completely determines the state of the working medium. The order of strokes in all thermodynamical cycles is crucial. It is reflected by
demanding at least two consecutive segment propagators to be non commuting $[{\cal U}_i,{\cal U}_j] \ne 0$.
Commuting propagators will lead to zero power.

\subsection{Limit Cycle} 

An important characteristic of the refrigerator is its limit cycle \cite{k201}.
All such devices, after some transient time settle to a steady state operational mode, the limit cycle.
The state of the working medium at the limit cycle becomes an invariant  of the propagator ${\cal U}_{global}$, 
with eigenvalue $\lambda_1=1$.  
The other eigenvalues of the propagator characterize the rate of approach to the limit cycle.
They obey the property $\lambda_{j+1} < \lambda_j < 1$.  
$\lambda_2$ determines a final exponential rate of approach to the limit cycle. 
The propagator after $n$ cycles is dominated by   $\lambda_2^n \approx e ^{- \gamma t}$,
where $\gamma = - \log(\lambda_2)/\tau_{cyc}$ and $\tau_{cyc}$ is the cycle time.

The limit cycle is determined by  external parameters defining the cycle. 
The invariant state is the analogue of the  thermodynamical equilibrium state. 
In the limit of very long times on each segment the limit cycle is completely determined by
the equilibrium conditions on the hot and cold \em isomagnetic \rm segments.
When there is insufficient time allocation to equilibrate, the limit cycle is constrained geometrically.
All the system observables become periodic. As a result the cycle trajectory in its parameter space
becomes fixed which also sets the performance.

The relation between the global constraint and the refrigerators performance will be demonstrated in an example
of a solvable working medium.

\section{The system }

The system  working fluid in this example is composed of pairs of coupled spins. 
The cycle parameter space is constructed from an operator vector space which is closed to the dynamics of all segments of the cycle. 
This vector space is sufficient to define the state of the working medium at all times.
On this vector space the cycle propagator is defined.

\subsection{ The Hamiltonian}

The Hamiltonian of the working fluid has the structure:
\begin{equation}
\Op H = \Op H_{int} + \Op H_{ext}(t)    
\label{eq:hamil}  
\end{equation} 
and in  general $ [\Op H_{int} ,\Op H_{ext}(t)] \ne 0$. 
The external Hamiltonian is chosen to be:
\begin{equation}
{\Op H}_{ext} ~~=~~\frac {1}{2} \omega(t)
\left({\boldsymbol{\mathrm{\hat{\sigma}}}}_z^1
\otimes {\bf \hat I^2}
+
{\bf \hat I^1} \otimes {{{{\boldsymbol
{ \mathrm {  \sigma}}}}}_z^2}
\right)~\equiv~\omega(t) {\Op B_1}
\label{extham}
\end{equation}
where $\omega(t)$ represents the external control field.
The uncontrolled, internal Hamiltonian is defined as ($\hbar=1$):
\begin{equation} 
{\Op H}_{int} ~~=~~\frac {1} {2} 
 J \left({ {\boldsymbol{\mathrm{\hat
{\sigma}}}}_x^1} \otimes 
{ {\boldsymbol{\mathrm{\hat{\sigma}}}}_x^2} -
{{\boldsymbol{\mathrm{\hat{\sigma}}}}_y^1}
\otimes {\boldsymbol
{\mathrm{\hat{\sigma}}}}_y^2 ~
  \right) ~\equiv~J {\Op B_2} 
\label{workingf}
\end{equation} 
where ${{\boldsymbol{\mathrm{\hat{\sigma}}}}}$
represents the Pauli operators, and $J$ scales the strength of the 
inter-particle interaction, which is assumed to be constant, for a given pump.

The total Hamiltonian becomes:
\begin{equation}
{ \Op H } ~~=~~\omega(t) {\Op B_{1}}+\rm J {\Op B_{2}}
\label{finalham}
\end{equation}
The eigenvalues of $\Op H$ are 
$ \epsilon_1= - {\Omega}(t) ,~ \epsilon_{2/3}=0,~ \epsilon_4= {\Omega}(t) $ 
where $\Omega(t)=\sqrt{\omega(t)^2+J^2}$. $\Omega(t)$ is the instantaneous  energy scale. 
 
The algebra of operators becomes closed by defining $\Op B_3$:
\begin{equation} 
[ {\Op B_1}, {\Op B_2}] \equiv  2 i {\Op B_3}~~=~~\frac{1}{2}  
\left({ {\boldsymbol{\mathrm{\hat
{\sigma}}}}_y^1} \otimes { {\boldsymbol
{\mathrm{\hat{\sigma}}}}_x^2} +
{{\boldsymbol{\mathrm{\hat{\sigma}}}}_x^1}
\otimes {\boldsymbol
{\mathrm{\hat{\sigma}}}}_y^2 ~~~  \right)
\label{Bthree}
\end{equation}

A vector space of operators which allows complete reconstruction of the density operator for the whole cycle
requires additional operators defined as:
${ {\Op B_4}= \frac{1}{2}({\Op I}^1 \otimes {\Op \sigma}_z^2 -{\Op I}^2
 \otimes {\Op \sigma}_z^1)}$ and ${{\Op B_5}={\Op \sigma}_z^1 \otimes {\Op \sigma}_z^2 }$.
The set $({\Op B_1}, {\Op B_2},{\Op B_3},{\Op B_4},{\Op B_5})$ constitutes  stationary orthogonal vector space
of operators with the scalar product defined as $ (\Op A \cdot \Op B) \equiv tr\{ \Op A^{\dagger} \Op B \}$.
We will show that it is sufficient to reconstruct the dynamics on all segments.

\subsection{Time dependent (linear) combination of the stationary algebra}

A  thermodynamically inspired set of observables is a minimum set of 
operators which completely defines the state of the working 
medium when it reaches the limit cycle. 
The set is initiated from the energy $\Op H$ and additional operators are added 
which are dynamically coupled to the energy.
This set is formed from a linear combinations of the stationary 
closed set ~$ \{ \Op B \} $ of  operators: 
\begin{equation}
{\Op H} ={\Large \omega}{ (t)} {\Op B_1 } +{\Large J} {\Op B_2 },
~
 {\Op L} ={\Large J} {\Op B_1 } - {\Large \omega}{(t)} {\Op B_2 }, 
~
 {\Op C}  = {\Large \Omega(t)} {\Op B_3 }
\label{defnew}   
\end{equation}
The three operators defined in Eq. (\ref{defnew}) form a closed Lie algebra. 
 
The additional two operators $\Op L$ and $\Op C$ are required  to uniquely define the diagonal part of 
the state $\Op \rho$ in the energy representation:

${\Op V} = {\Omega} {\Op B_4}$ and
 
${\Op D} = {\Omega} {\Op B_5}$. The operators $\Op V$ and $\Op D$ commute with $\Op H$. 
In this operator base the state $\Op \rho$ can be expanded as:
\begin{equation}
\Op \rho ~=~
\frac{1}{4} {\Op I} + \frac{1}
{ \Omega} \left( \langle \Op H \rangle \Op H +
\langle \Op L \rangle \Op L 
+\langle \Op C \rangle \Op C  + \langle \Op V 
\rangle \Op V +\langle \Op D \rangle \Op D  \right)
\label{eq:rho}
\end{equation}
The equilibrium value of $\langle \Op V \rangle$ is  zero as well as all the remaining operators
of the Hilbert space which are not generated from $\Op H$. 
Therefore the state $\Op \rho $ of the cycle can be reconstructed by only four expectation values:
$E = \langle \Op H \rangle $,  $L = \langle \Op L \rangle $, $C = 
\langle \Op C \rangle $ , $D = \langle \Op D \rangle $ and the identity $\Op I$.
In the energy representation the state $\Op \rho$ has the explicit form:
\begin{eqnarray}
\Op \rho_e = \frac{1}{4} \left(
\begin{array}{cccc}
1+ \frac{1}{\Omega}(D -2E)&0&0&\frac{2}{\Omega}(L+iC)\\
0&1- \frac{1}{\Omega}D &0&0\\
0&0&1- \frac{1}{\Omega}D &0\\
\frac{2}{\Omega}(L-iC)&0&0&1+ \frac{1}{\Omega}(D+2E)
\end{array}
\right)
\label{eq:rhoe} 
\end{eqnarray}
where $\Op \rho_e$ is the density operator in the instantaneous energy base.

\section{The equations of motions and their solutions; the propagators.}
\label{eqmotion}

The dynamics of an operator $ {\Op A}$  is generated by a completely positive 
map, a Liouville superoperator, $ {\cal L}$ \cite{alicki87}.  In the Heisenberg picture it becomes
\begin{equation}
\frac {d {\Op A}}{dt}~~=~~ i[{\Op H}, 
{\Op A}]+ {\cal L}_{D}( {\Op A})
~+~ \frac{\partial {\Op A}}{\partial t}~~~.
\label{eq:heisenberg}
\end{equation}    
$ {\Op A}$ can be explicitly time dependent. ${\cal L}_D$ is the dissipative superoperator
leading to thermal equilibrium with the bath temperature \cite{lindblad76}.

\subsection{The Propagators  on the {\em Isomagnets }}
\label{proisom}

The vector space defining the propagators is constructed from the operators 
$(\Op H,\Op L,\Op C , \Op D,\Op I)$.
The equations of motion on the {\em isomagnetic } thermalisation segments are \cite{k258}:
\begin{eqnarray}
\frac {d}{dt}
\left(
\begin{array}{c}
  {\Op H}  \\
  {\Op L} \\
  {\Op C} \\ 
  {\Op D} \\
  {\Op I} \\
\end{array}
\right)~~=~~ 
\left( 
\begin{array}{ccccc} 
-\Gamma & 0 & 0 & 0 & \Gamma E_{eq} \\
 0 & -\Gamma & -\Omega & 0 &  0 \\ 
 0 & \Omega &  -\Gamma & 0 &  0 \\ 
\frac{2}{\hbar \Omega} E_{eq} & 0 & 0 & - 2 \Gamma & 0 \\
 0 & 0 & 0 & 0 &  0 \\  
\end{array}
\right)
\left(
\begin{array}{c}
  {\Op H}  \\
  {\Op L} \\
  {\Op C} \\ 
  {\Op D} \\
  {\Op I} \\
\end{array}
\right)
\label{eq:eqiso}  
\end{eqnarray}  
where $\Gamma=k^++k^-$ ,  $k^+/k^-=\exp(\hbar \Omega/kT)$ and $E_{eq}=\hbar \Omega(k^+-k^-)/\Gamma$.

Using this set the propagator $ {\cal U}_{i}(\tau)$  on the  {\em isomagnets \rm} becomes \cite{k274};
\begin{eqnarray}
{\cal U}_i =
\left(
\begin{array}{ccccc}
e^{(-\Gamma_{i} \tau)}  & 0 & 0& 0  & {E}_{eq}
(1-e^{(-\Gamma_{i} \tau)} \\
0  &  K_i \cos(\Omega \tau)      & 
-K_i\sin(\Omega \tau) & 0 & 0  \\
0  &  K_i\sin(\Omega \tau) & K_i \cos(\Omega 
\tau) & 0 & 0 \\ 
\frac {1}{\Omega}(E_{eq}(e^{-\Gamma_{i}\tau}-
e^{-2\Gamma_{i}\tau}))   &  0 & 0 & e^{-2\Gamma_{i} \tau}  
&- \frac{E_{eq}^{2}}{\Omega}(e^{-\Gamma_{i} \tau}-1) \\
0  &  0   & 0  & 0 & 1  \\
\end{array}
\right)
\label{eq:propiso}  
\end{eqnarray} 
where $ K_i= e^{(-[\Gamma_{i}] \tau )}$ and $i=c,h$.
The periodic functions in Eq. (\ref{eq:propiso})  
mean that the isomagnetic segments are quantized.  
Whenever $\Omega \tau=2 \pi$, the two coupled equations
of $ L,C$  complete a period.

\subsection{Propagators on the {\em adiabats }}
\label{sec:subapproxadi} 

The equation of motion on the {\em adiabats} are:
\begin{eqnarray}
\frac {d}{\Omega dt}
\left(
\begin{array}{c}
  {\Op H}  \\
  {\Op L} \\ 
  {\Op C} \\
\end{array}
\right)
(t)~~=~~
\left( 
\begin{array}{ccc}
\frac{\dot \Omega}{\Omega^2} &  -\frac {J \dot \omega}{\Omega^3}  & 0 \\
\frac {J \dot \omega}{\Omega^3} & \frac{\dot \Omega}{\Omega^2}  & -1  \\
0  &  1 & \frac{\dot \Omega}{\Omega^2} \\
\end{array}
\right)
\left(
\begin{array}{c}
  {\Op H}  \\
  {\Op L} \\
  {\Op C} \\    
\end{array}
\right)
\label{eq:eqadi}  
\end{eqnarray} 
Closed form solutions for the  {\em adiabats} 
leading to quantized motion can be obtaind 
for a constant adiabatic parameter $\mu =  \frac {J \dot \omega}{\Omega^3} $.
$\mu \rightarrow 0$ is the adiabatic limit, then the system stays on the energy shell.
When $\mu$ increases coherences are generated due to non adiabatic transitions.

The exact solution for  the propagator on the {\em adiabats }, 
${\cal U}_{hc}$ for a constant adiabatic parameter $\mu$ is \cite{k274}:
\begin{eqnarray}
{\cal U}_{hc}~~=~~\frac{\Omega_c}{\Omega_h}
\left(
\begin{array}{ccccc}
\frac {1 + \mu^2 c}{q^2}   & -\frac{\mu s}{q}   & 
 \frac {\mu(1-c)}{q^2}&0&0   \\
 \frac {\mu s} { q}       & c    &- \frac {s}{q} &0&0  \\
 \frac {\mu (1-c)}{ q^2}  & \frac {s}{q} & 
\frac {\mu^2+c} {q^2}  &0&0\\
0&0&0&1&0\\
0&0&0&0&\frac{\Omega_h}{\Omega_c}\\
\end{array}
\right) ~~,
\label{eq:calprop}  
\end{eqnarray}   
where  $\Omega_{c/h}=\sqrt{J^2+\omega_{h/c}^2}$, $q= \sqrt{1 +\mu^2}$, $s=sin(q \Theta)$, 
 $c=cos(q \Theta)$. and $\Theta=J\frac{\arcsin(\frac{\omega_c}{\Omega_c})- 
\arcsin(\frac{\omega_h}{\Omega_h})}{\frac{\omega_c}{\Omega_c}- \frac{\omega_h}{\Omega_h}}$. 
The propagator ${\cal U}_{ch}$ has a similar form.

The propagator on the {\em adiabat} is periodic when $q \Theta= 2 \pi l$, $l$ integer,
then the propagator restores to the identity meaning there is no mixing between $\Op H, \Op L, \Op C$.
For half integer values of $l$, $l=1/2,3/2...$ there is maximum mixing.

\subsection{Two Time Scales }

When operating the engine, two time scales arise. The cycle time $\tau_{cyc}$ and the
internal time scale of the working fluid, determined by
the energy spectrum $\tau_{int} ={ \frac{2 \pi}{\Omega}} $.
In relation to these time scales the  cycles can be classified as either regular or sudden:
The {\em regular \rm} set, where ${ \tau_{cyc} }~\gg~\tau_{int} $ and
the {\em sudden \rm} set, where  ${ \tau_{cyc} }~\ll~\tau_{int} $.
In the limit of very long time the regular cycles become adiabatic meaning that at all segments
the state of the working medium is completely defined by its energy expectation value.

In contrast the sudden cycles are functions of the expectation values of all the operators 
defining the vector space. The deviation from regular cycles is characterized by  coherence:
The coherence measure is defined  as a square distance from a state diagonal in energy \cite{k108,girolami2014observable}:
\begin{equation}
\tilde {\cal C} = tr \{ (\Op \rho - \Op \rho_{ed})^2\}~,
\label{eq:coherence}  
\end{equation}
where $\Op \rho_{ed}$ is the diagonal stationary part of the density operator 
in the energy frame. 
From Eq. (\ref{eq:rhoe}),~ the coherence measure becomes 
$\tilde {\cal C}= \frac {L^2+C^2} { \Omega^2}$. 

The quantum character of the cycles is also elucidated by comparing
the von Neumann entropy ${\cal S}_{VN}= - tr \{ \Op \rho \ln \Op \rho \}$ to the energy entropy ${\cal S}_E = - tr \{ \Op \rho_{ed} \ln \Op \rho_{ed} \}$ where ${\cal S}_{VN} \le {\cal S}_E$.
The difference ${\cal S}_E- {\cal S}_{VN} = Tr \{ \Op \rho (\ln \Op \rho- \ln \Op \rho_{ed} \}$ is the conditional distance $D(\Op \rho|| \Op \rho_{ed})$ between the state $\Op \rho$ to a state on the energy shell with the same energy.
The term $\beta_{ed}D(\Op \rho|| \Op \rho_{ed})$ characterises the irreversible work required to generate  coherence
from  the diagonal state $\Op \rho_{ed} $ \cite{plastina2014}.

\begin{figure}[tb]
\vspace{2.2cm} 
\center{\includegraphics[height=7.5cm]{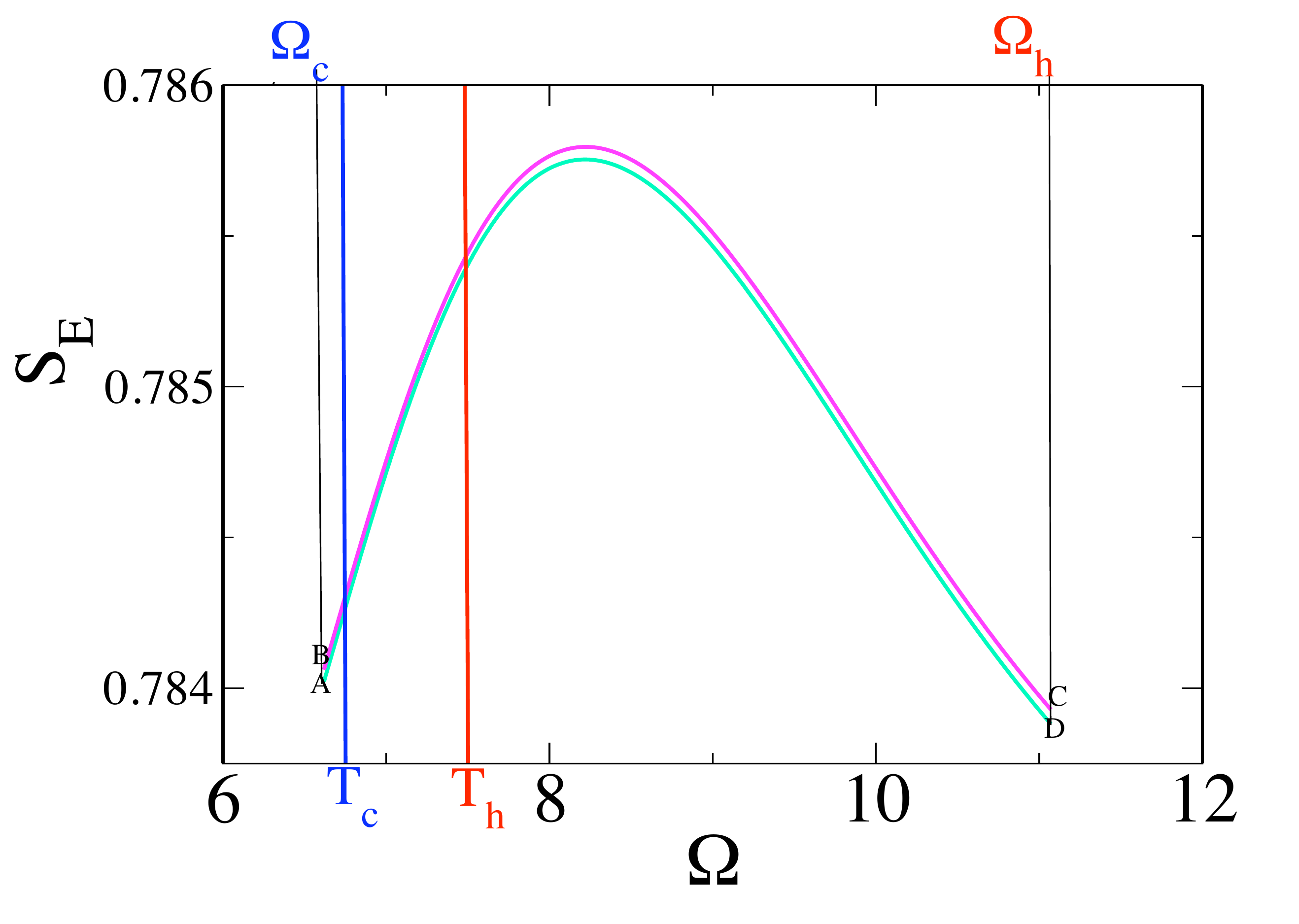}} 
\vspace{0.2cm} 
\caption{ Energy entropy vs internal energy scale $(\Omega,S_E)$ for a typical {\em sudden} cycle. The cold and hot isotherms are also indicated.
For this cycle the two isomagnetic segments (A,B) and (C,D) are very short reflected in small changes
in ${\cal S}_E$. For this cycle the expansion and compression {\em adiabats} are very close.
The cycle parameters:
$  J=1.25,T_h=4,~T_c=3.6,\omega_c=6.5,~\omega_h=11,~~
\kappa_h^{\downarrow}=0.36,~ \kappa_c^{\downarrow}=0.0656$.
The times on the different segments are:
$\tau_{h,hc,c,ch}=0.00045,0.466,0.03282,0.466$. The cooling power is 
$~Q_c/\tau=1.45*10^{-4}$. The Von Neumann entropy $\Delta S_{VN}=4.8*10^{-5}$
} 
\label{fig:1}   
\end{figure}

\section{The relation between cycle geometry and performance  of the {\em sudden} cycles}
\label{charprop}

The performance of the sudden refrigerators is extremely sensitive the the cycle parameters \cite{k274}.
These parameteres determine the geometry of the cycle and with it its performance.
In order to carry out a systematic study we maintain all parameters fixed except the cycle time $\tau_{cyc}$
which is varied. The ratio $\frac{\Omega_c}{\Omega_h}$ and  $ \frac{T_c}{T_h}$ are constant and maintain 
$\frac{\Omega_c}{\Omega_h}> \frac{T_c}{T_h}$ which is a necessary condition for a refrigerator. Nevertheless,
for certain cycle times the device operates as a short circuit converting work into heat on both the hot and cold baths.

We chose to start the study from the family of sudden cycles 
where $\omega_h , \omega_c \gg J$ and $\tau_{adi} \sim 1$.  These cycles termed
3(b) of ref \cite{k274}  have quantum character revealed by an approximate global propagator: ${\cal U}_{global}^{3b}$.
Additional cycles were generated by systematically decreasing  the total cycle time $\tau_{cyc}$. 
The fraction time allocation for each segment was kept constant
therefore: $\tau_{hc} = \tau f_{hc}, ~ \tau_c=\tau f_c,~\tau_{ch}=\tau f_{ch},~\tau_h=\tau f_h$ 
where $f_{hc}=0.48277,~f_c=0.0340,~f_{ch}=0.48277,~f_h=0.000466~$.

\begin{figure}[tb]
\vspace{1.8cm} 
\center{\includegraphics[height=7.5cm]{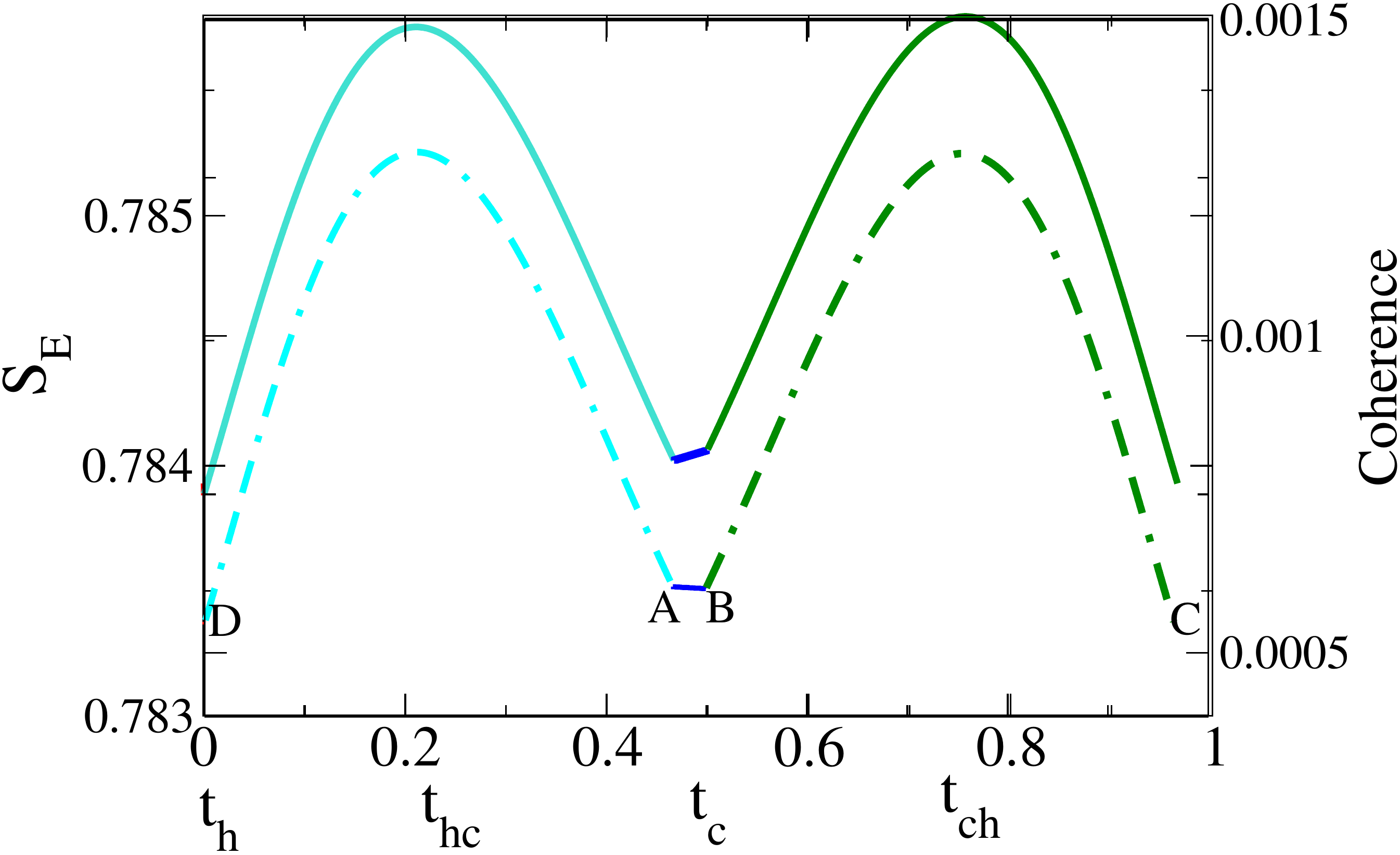}}
\vspace{0.1cm} 
\caption{Energy entropy $S_E$ ( solid left scale ) and Coherence ( dashed, right scale ) as a function 
of time along the cycle.~  Data are of cycle (b) in Fig. \ref{fig:fourconc} }
\label{fig:noTdyn}  
\end{figure}

Figure \ref{fig:1} shows a typical cycle in the $(\Omega,{\cal S}_E)$ plane.
The state of these cycles is never close to adiabatic conditions. This is evident by the large coherence at all times
reflected in the coherence measure Cf. Fig \ref{fig:noTdyn}. In addition   
a  large gap between the energy entropy and the von Neumann entropy characterises the entire cycle Cf. Fig. \ref{fig:fourconc}.

 The cycle trajectories for all intermediate  times were obtained by direct numerical integration of Eq. (\ref{eq:heisenberg}) sequentially for all branches, resulting in the expectation values  of the vector space of operators 
 $\Op H \Op L \Op C \Op D$ for all times. The limit cycle vector was obtained by propagating for many periods
 until convergence was obtained.
 The exact propagator ${\cal U}_{global}$ which is the product of the segment propagators was represented
 numerically. Eigenvectors and eigenvalues  were obtained for point A. The eigenvector corresponding to 
 eigenvalue $\lambda_1=1$ is the cycle invariant vector.
 The limit cycle invariant vector obtained by the two methods was compared converging to at least $10^{-6}$.
 In addition the approximate propagator ${\cal U}_{global}^{3b}$ was diagonalized and its
 eigenvalues compared to the exact eigenvalues of ${\cal U}_{global}$, for different cycle times (Cf. Fig. \ref{fig:eig} ).

\subsection{Geometrical evolution of the  {\em Sudden} Cycles} 
\label{subsec:manyshapes}
  
\begin{figure}[tb]
\vspace{2.2cm} 
\center{\includegraphics[height=15.5cm]{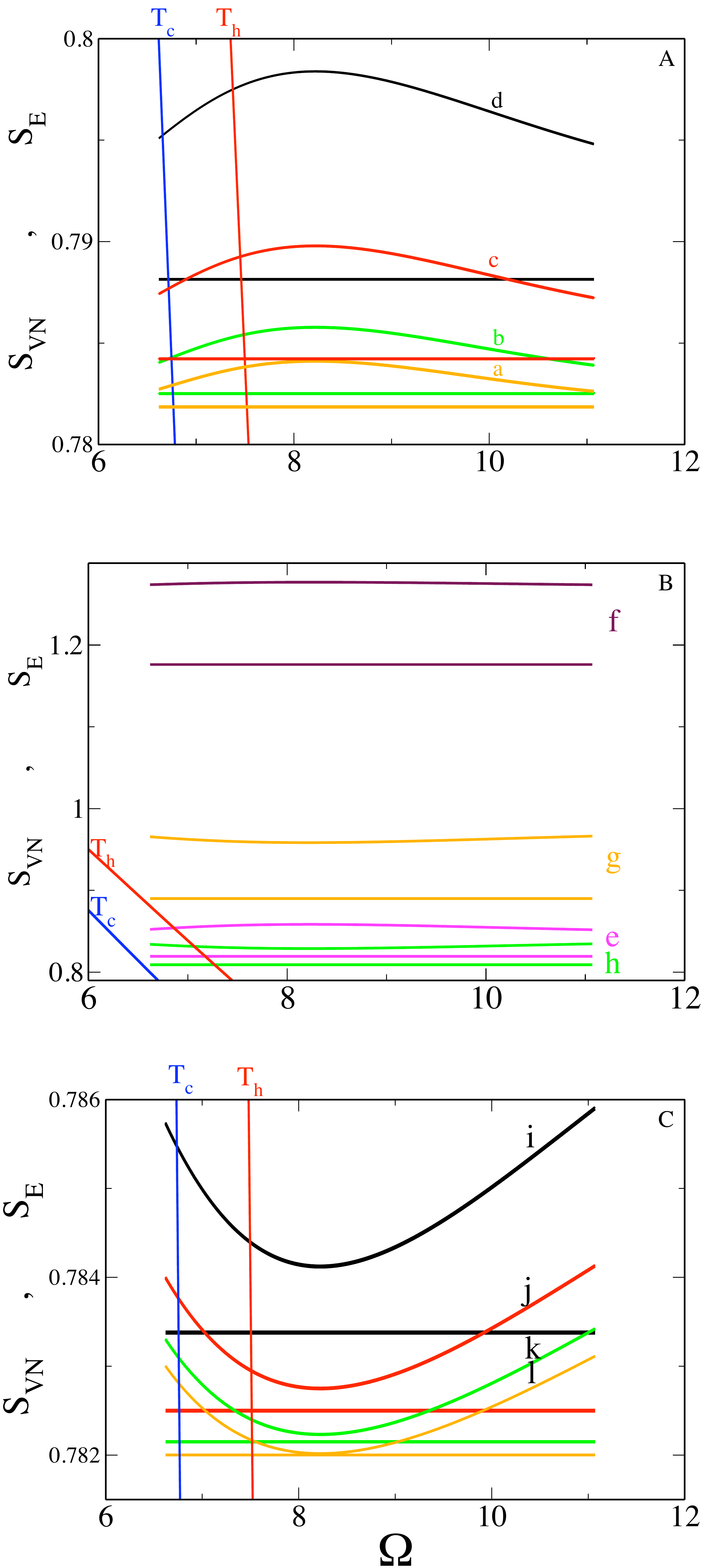}} 
\vspace{0.2cm} 
\caption{ The entropy vs cycle energy scale for a family of refrigerator cycles with decreasing cycle time plotted both in the planes of $(\Omega,S_{VN})$  and $(\Omega,S_E)$ . 
The von Neumann entropy $(\Omega,S_{VN})$ seen as  horizontal lines in the figure are almost constant
in the scale of the plots. Notice the change from convex to concave upon decreasing cycle time.
The common data of the cycle family is:
$  J=1.25,T_h=4,~T_c=3.6,\omega_c=6.5,~\omega_h=11,~~
\kappa_h^{-}=0.36,~ \kappa_c^{-}=0.0656$
The cycle times for figure {\bf A}
a,b,c,d is: $\tau_a=1.013534,~\tau_b=0.96527,\tau_c=0.912180, \tau_d=0.868743$.
The corresponding cooling rates $Q_c/\tau$  are: $1.57*10^{-4},~ 1.445*10^{-4},~
 1.133*10^{-4},~ 4.142*10^{-5}$
The cycle times for figure {\bf B}, e,f,g,h is: 
$\tau_e=0.81683,~\tau_f=0.772216,~\tau_g=0.735444,~\tau_h=0.700423$.
The corresponding $Q_c/\tau$ are:  $-5.43*10^{-4},~ -8.2*10^{-3},~ -1.93*10^{-3},~
-3.47*10^{-4}$ 
The cycle times for figure {\bf C}, i,j,k,l is: 
$\tau_i=0.482635,~\tau_j=0.36197625,~\tau_k=0.2413175, ~\tau_l=0.12065875$.
The corresponding $Q_c/\tau$ are: $1.29*10^{-4},~ 1.45*10^{-4},~ 1.513*10^{-4},~
1.541*10^{-4}$} 
\label{fig:fourconc}   
\end{figure}

\begin{figure}[tb]
\vspace{1.2cm} 
\center{\includegraphics[height=5.5cm]{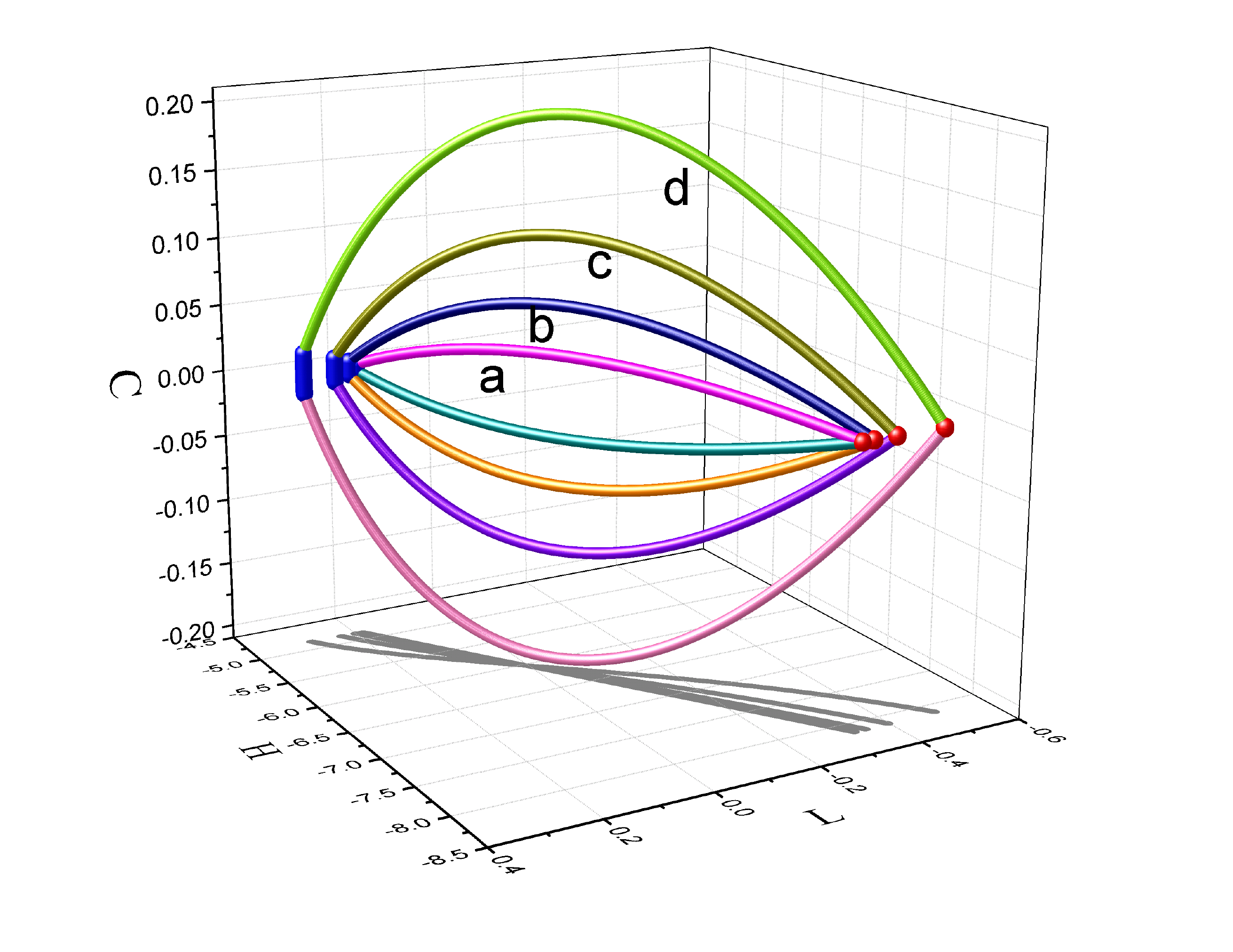}}
\center{\includegraphics[height=5.5cm]{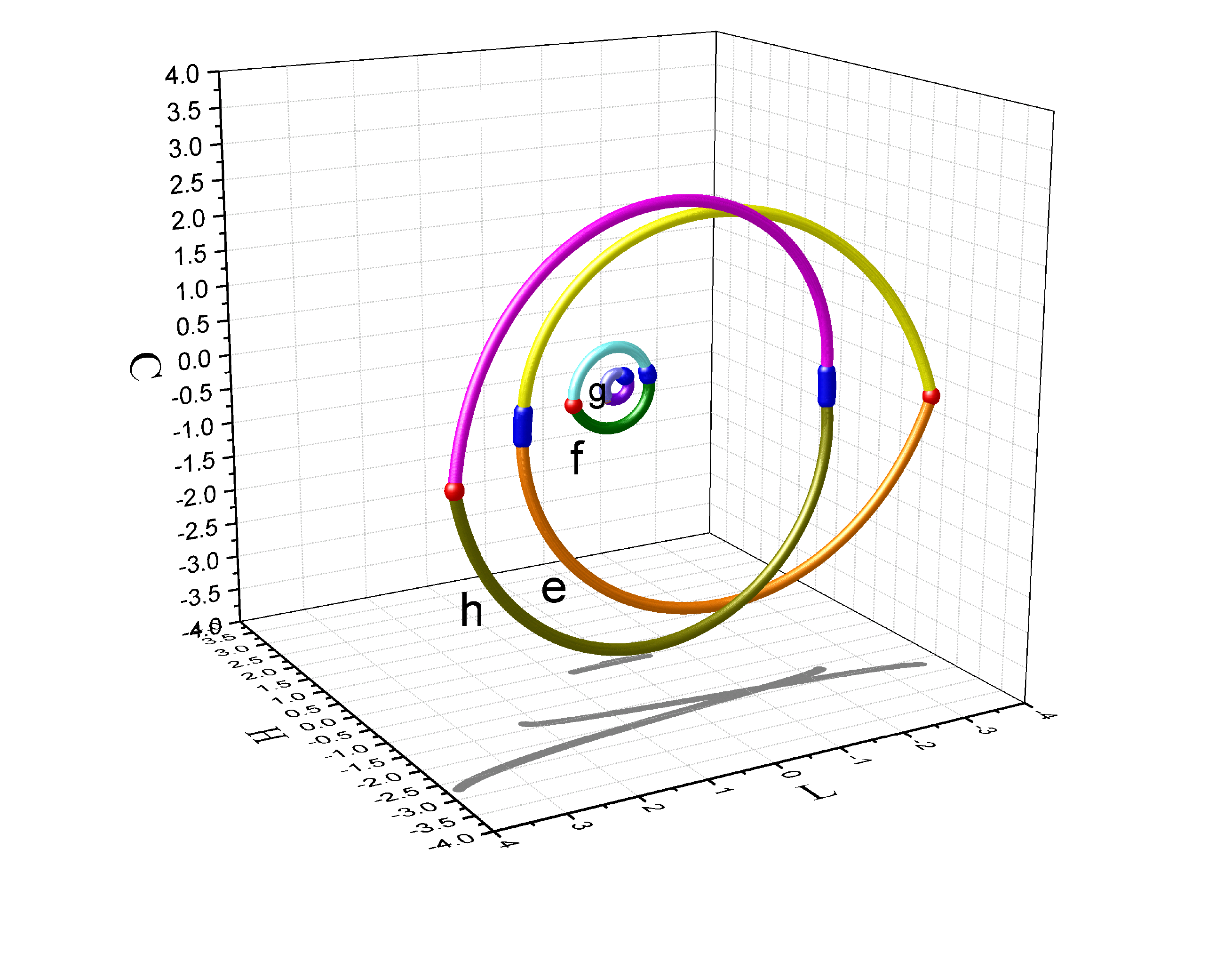}} 
\center{\includegraphics[height=5.5cm]{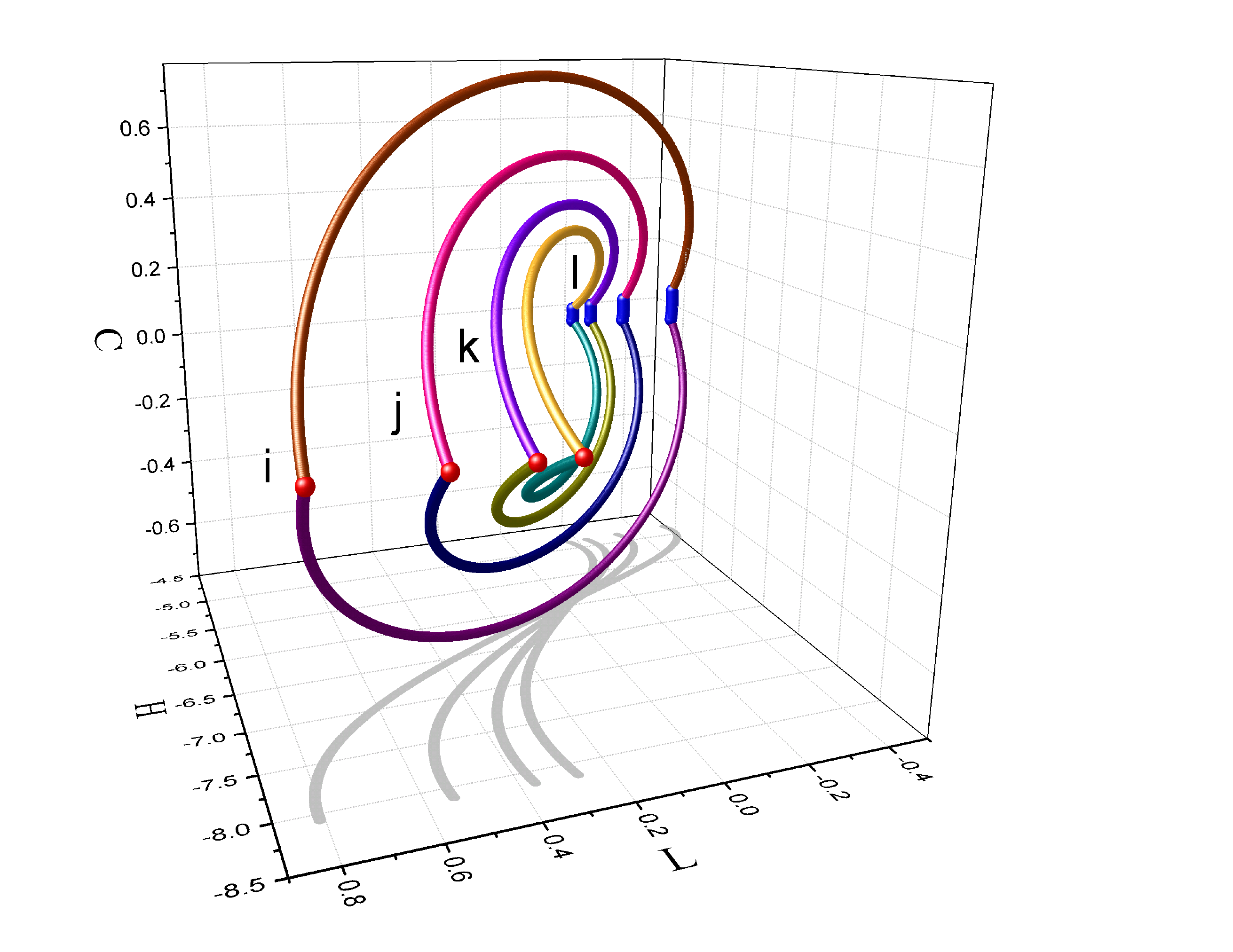}} 
%\center{\includegraphics[height=15.5cm]{combtraj1.eps}} 
\vspace{0.2cm} 
\caption{The trajectories in the $(H,L,C)$ vector space in descending order of cycle time. 
The blue sections represent cold {\em isochores} the red sections hot {\em isochores}. 
Most of the cycle trajectories is occupied by the {\em adiabats}. 
Top and bottom panel shows refrigerator cycles. Notice that the change in derivative $\frac{d H}{dL}$ 
such that the cycles are almost orthogonal. The middle panel shows non refrigerator cycles where
the transition between the two types of refrigerators occurs. 
The parameters correspond to  Fig. (\ref{fig:fourconc}). Each panel is normalised differently.} 
\label{fig:combtraj}    
\end{figure}

Figs. \ref{fig:fourconc} and \ref{fig:combtraj}, show refrigerator cycles of different shapes with decreasing cycle times. 
The cycles are represented both in $(\Omega,{\cal S}_{VN}),~~(\Omega,{\cal S}_E)$ planes, and in $(H,L,C)$ vector space. 
We follow the evolution of cycle shapes from the largest cycle times in Figs. \ref{fig:fourconc} A, representing
concave refrigerator cycles.
As the cycle time becomes shorter  the entropies and the gap between entropies ${\cal S}_E$ and  ${\cal S}_{VN}$ increases.
It should be noted that on the scale of the graphs the compression and expansion 
{\em adiabats} seem to merge.
This observation suggests a trend that decreasing cycle time leads to larger coherence.
When the cycle time is further decreased, the cycles cease to be refrigerators Cf. Fig. ~\ref{fig:fourconc} B.

Decreasing the cycle time leads to more concave cycles in the $(\Omega,{\cal S}_E)$ plane
and a larger incursion in the $L$ and $C$ direction Cf. Fig. \ref{fig:combtraj}. In addition  the correlation observable $C$ changes sign between the compression and expansion {\em adiabatic}. On both {\em adiabats} the derivative $\frac{dH}{dL} < 0$.
           
\begin{figure}[tb]
\vspace{1.2cm} 
\center{\includegraphics[height=6cm]{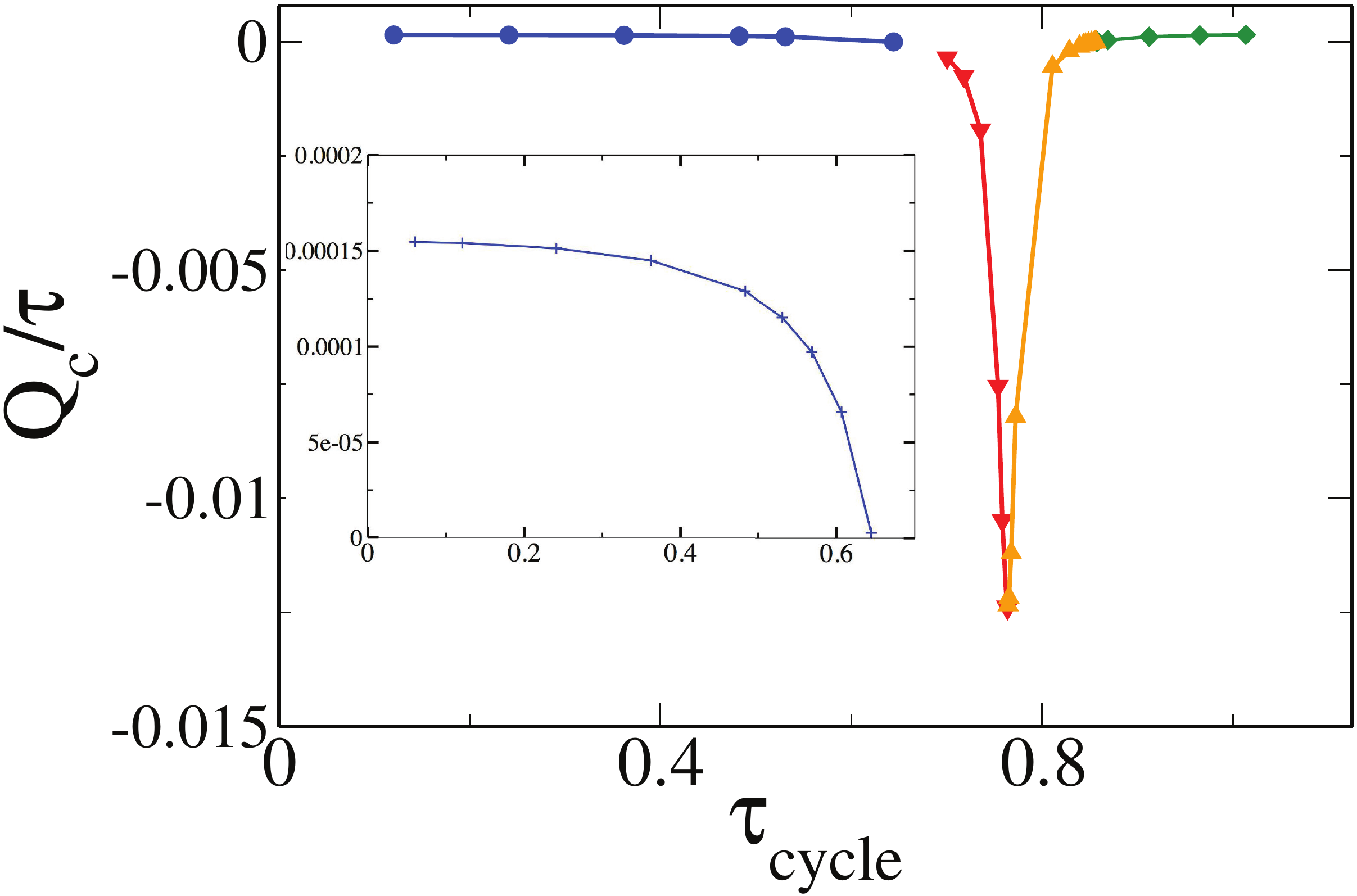}} 
\vspace{0.3cm}
\center{\includegraphics[height=6.05cm]{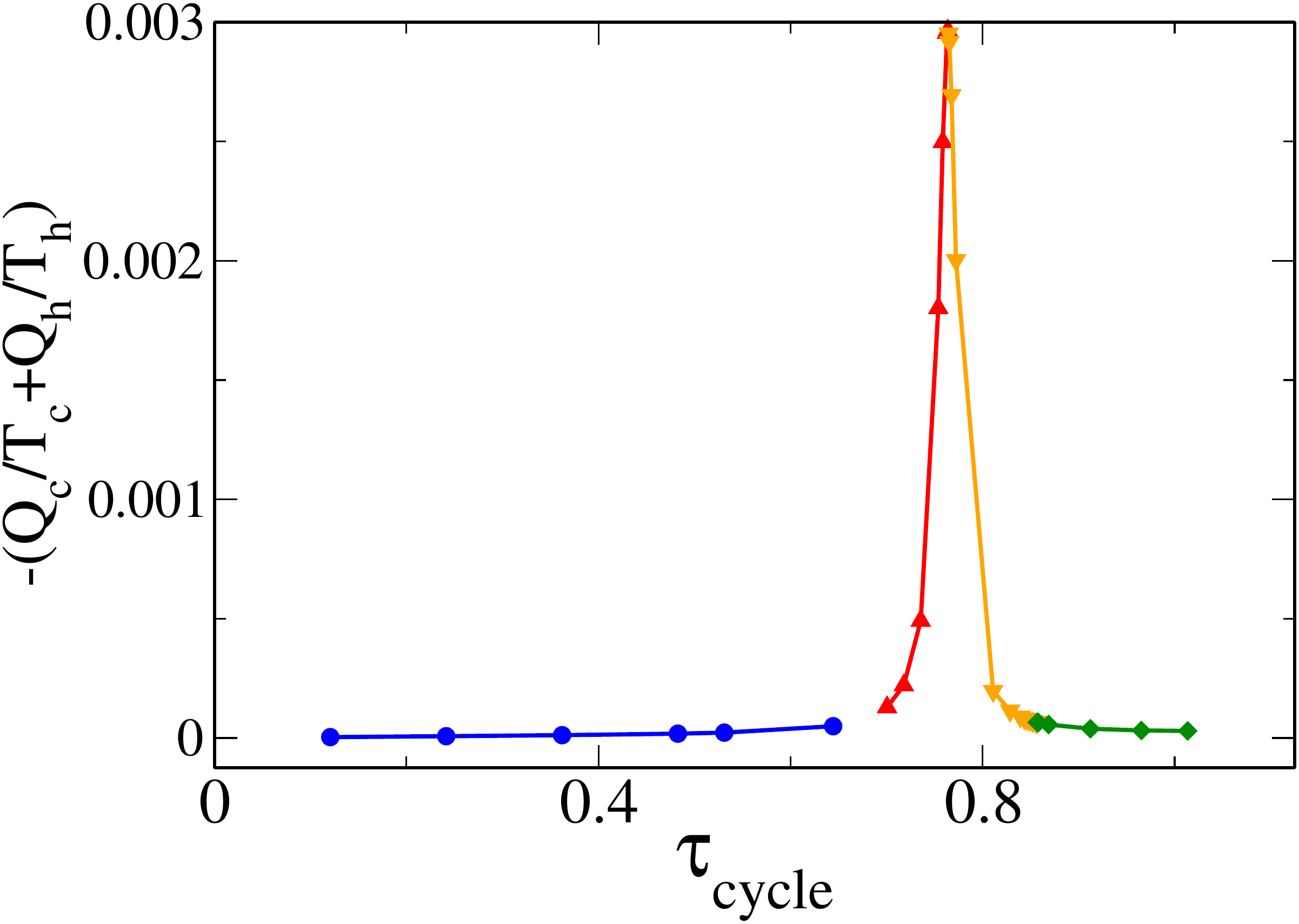}}
\vspace{0.1cm} 
\caption{ Top:  cooling power as a function of cycle time. The inset shows the ultrashort refrigerators reaching an asymptotic cooling power when the cycle time approaches zero. This limit corresponds to the universal region of small action.
The blue and green segments correspond to refrigerator cycles $\frac{Q_c}{\tau} >0$ and the orange and red to
non-refrigerator cycles $\frac{Q_c}{\tau} <0$. Bottom: Entropy $\Delta {\cal S}_u $ generated in the baths vs cycle time.
The common data: $ J=1.25,T_h=4,T_c=3.6,\omega_c=6.5,\omega_h=11.,
\kappa_h^{-}=0.36,~\kappa_c^{-}=0.0656$.  } 
\label{fig:typical2}      
\end{figure}

\begin{figure}[tb]
\vspace{1.2cm} 
\center{\includegraphics[height=7cm]{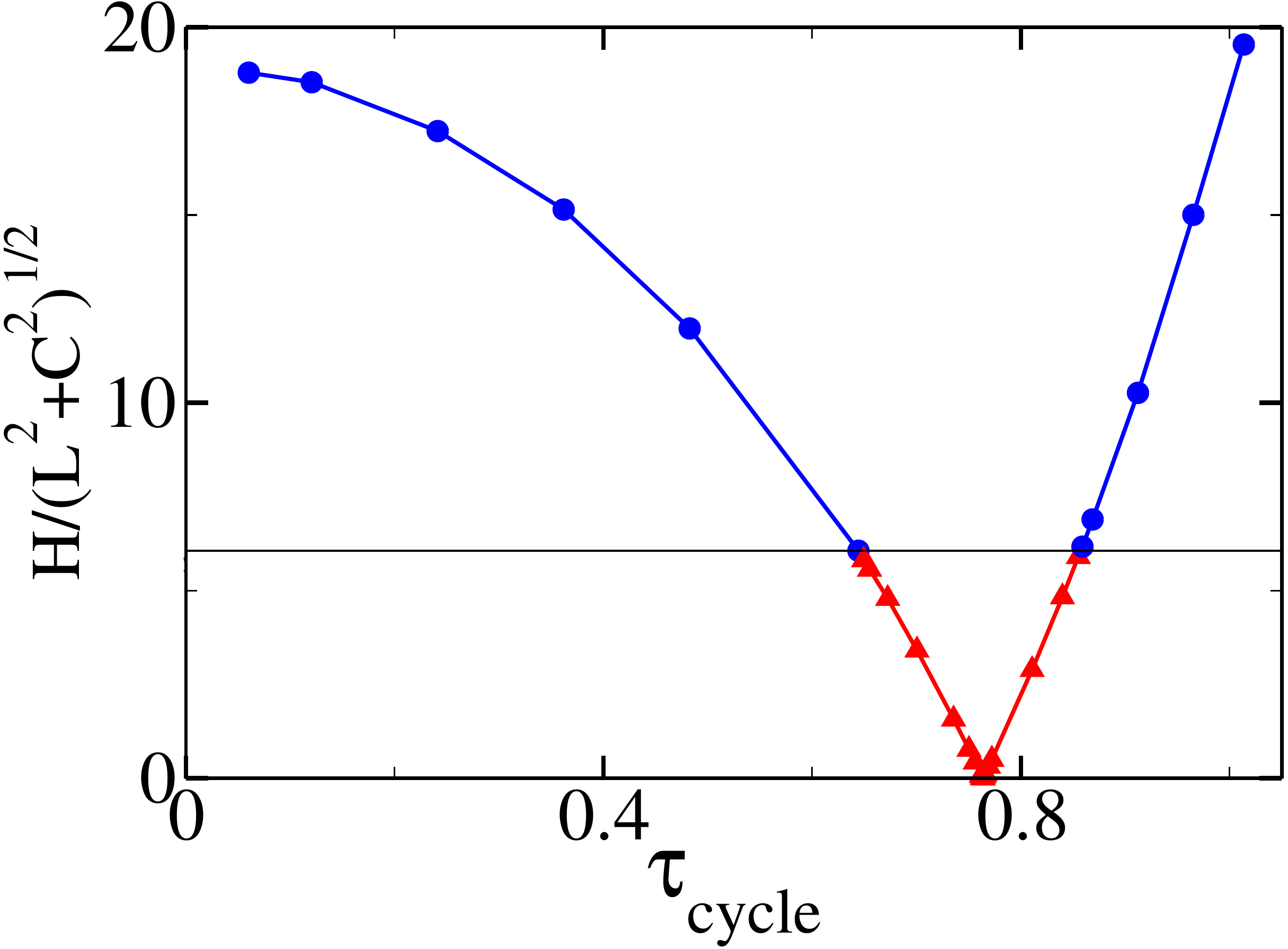}} 
\vspace{0.1cm} 
\caption{The ratio between the energy and coherence $\frac{H}{\sqrt{L^2+C^2}}$ as function of $\tau_{cycle}$
demonstrating a sharp transition from refrigerators(blue cycles)
to non refrigerators (red triangles up) and back into refrigerators.
The other data are as in Fig. \ref{fig:typical2}.  The calculations correspond to point A in the cycle
(the transition between the {\em adiabat} and cold {\em isochore}).
}
\label{fig:HperCOH}      
\end{figure}

\begin{figure}[tb]
\vspace{1.2cm} 
\center{\includegraphics[height=7cm]{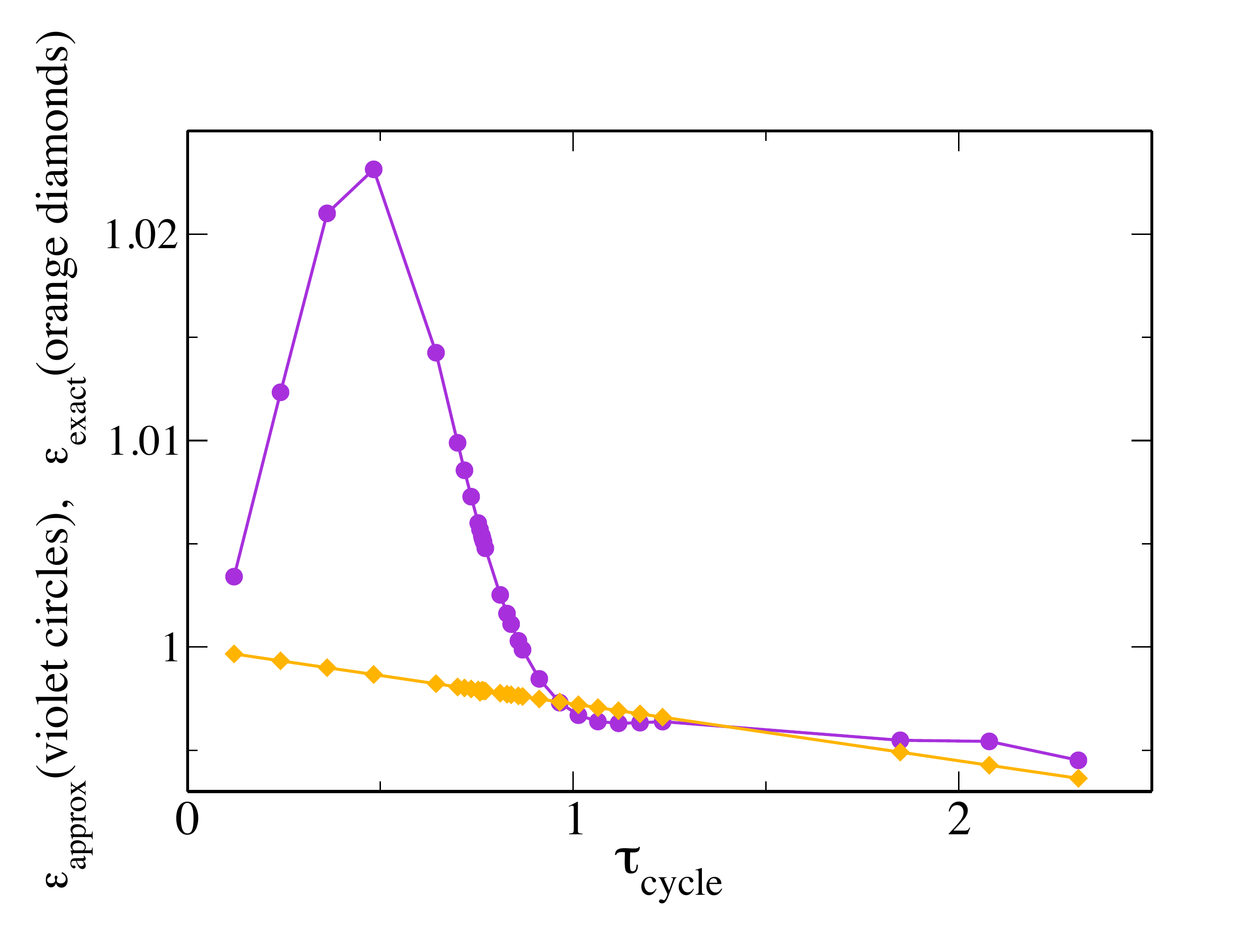}} 
\vspace{0.1cm} 
\caption{The eigenvalue $\lambda_2$ as a function of cycle time. The orange points represent the 
eigenvalus of the exact propagator ${\cal U}_{global}$. 
The purple points are $\lambda_2$ of  approaximate ${\cal U}_{global}^{3b}$.  The cycle time
where the approximate eigenvalues are degenerate is the transition point from refrigerator to non-refrigerator.}
\label{fig:eig}      
\end{figure}
  
A further decrease in cycle time leads to an additional increase in entropy 
as well as the entropy gap between ${\cal S}_{VN}$ and ${\cal S}_E$.  
This is accompanied by  vanishing of the cooling power ${\cal Q}_c/\tau <0$. 
The cycle dissipates the power to both the hot and cold baths, a short circuit. 
Figure \ref{fig:typical2} shows the cooling power and the entropy generation as a function of cycle time.

The transition point in cycle period between a refrigerator and short circut was found to match  a point of degeneracy in the eigenvalues
and eigenvectors of the approximate propagator ${\cal U}_{global}^{3b}$ i.e. 
$\lambda_2 \rightarrow \lambda_1 =1$.
This degeneracy was absent in the exact propagator indicating that the approximation loses its validity beyond the transition point.
Cf. Fig \ref{fig:eig}.
The second eigenvalue of the exact propagator is monotonically decreasing with cycle time reflecting the increasing dissipation
due to more time allocated to the hot and cold {\em isochors}.

A maximum in entropy production is located in the region of non-refrigerators Cf. Fig. \ref{fig:typical2}.
This point is also a minimum in negative cooling power. This is highly understandable since
$Q_c>0$ and  $Q_h>0$ so that $\Delta S_u= Q_h/T_h+Q_c/T_c$ becomes large since both terms are positive 
in contrast to engines and refrigerators.

\begin{figure}[tb]
\vspace{2.2cm} 
\center{\includegraphics[height=8.5cm]{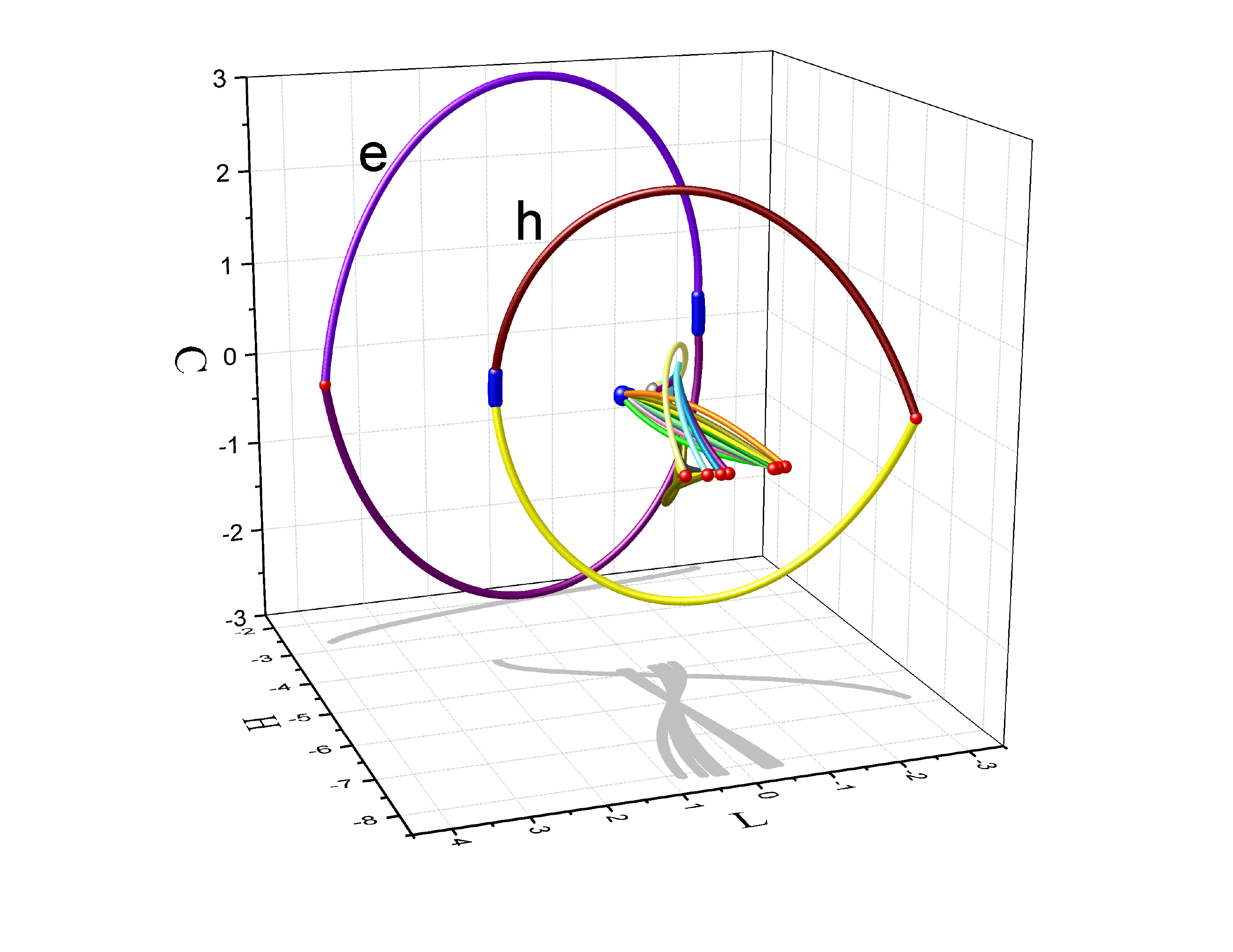}} 
\vspace{0.2cm} 
\caption{All refrigerator and non refrigerator cycles in the  $(H,L,C)$ space the same as in figure \ref{fig:combtraj}.
The short circuit cycles indicated by $e,h$ occupy a much larger volume in the $H,L,C$ space reflecting the significantly
larger entropy generation associated with large values of $L$ and $C$. The refrigerator cycles are concentrated
in a smaller volume. The same normalisation is used for all cycles. The vector space of the $(H,L,C)$ set defines a scalar product between cycles. The concave and convex cycles are approximately orthogonal.} 
\label{fig:alltraj}   
\end{figure}

Further decreasing the cycle period for the set of non refrigerators results in a dramatic geometrical
change of  cycles. In the plane of $ (\Omega, S_E)$,  the cycles change from concave to convex. 
In the the three dimensional $H,L,C$ space the shape changes from an eye like trajectory to a squashed set
Cf. Fig. \ref{fig:combtraj} C. 
The transition point from concave to convex cycles is also a maximum point in the entropy production 
and a minimum in the cooling power ${\cal Q}_c/\tau$, Cf. Fig. \ref{fig:typical2}. We also
find drastic change in the character of the cycles going through the maximum. $H$ becomes small and $\frac{dH}{dL}$ becomes almost zero.
The cycle closest to the extreme point (cycle g) is almost a perfect circle in the $L,C$ plane.
Figure \ref{fig:HperCOH} shows that the ratio of energy to coherence is minimum at the transition point.

Beyond this point decreasing the cycle time the convex character increases and another transition point appears.
The cycles act as a refrigerators again and the derivative $\frac{d H}{d L}$ becomes positive. 
Notice the insert in Figure \ref{fig:typical2} demonstrating that the cooling power reaches a non zero asymptotic value
when $\tau_{cyc} \rightarrow 0$. This is also true for the ratio between energy and coherence in Figure  \ref{fig:HperCOH}.

\section{Rationalising the geometrical and perfomance: points of transitions.}

The phenomena observed is that
with decreasing cycle time the family of cycles goes through  two transition points from 
refrigerator to short circuit cycles and back to refrigerators cycles. In between these points the  topology 
changes in an extreme point of entropy production.

We first address the geometrical change. The family of cycles studied is dominated by time allocation
to the {\em adiabats}. The cycle $g$ is almost circular and the coherence dominates. This cycle is the 
result of quantization on the {\em adiabats} for half integer $l=\frac{1}{2}$ on each of the adiabtic segments, Eq. (\ref{eq:calprop}).
In this case  $c = \cos(q \Theta)=-1$ and $s=\sin(q \Theta)=0$.
The quantization condition becomes:
\begin{equation}
\tau_a = \bar K \sqrt{\left(\frac{ 2 \pi l} {\Phi}\right)^2-1}
\label{eq:quanti}
\end{equation}
where $\bar K=\frac{1}{J}\left(\omega_c/\Omega_c-\omega_h/\Omega_h\right)$ and $\Phi=\arcsin(\frac{\omega_c}{\Omega_c})- 
\arcsin(\frac{\omega_h}{\Omega_h})$.
This point $l=\frac{1}{2}$ has maximum coherence leading to $\tau_a=0.38$ which is half the cycle time since
the time allocted to the isochores is negligible. 
The topological change is generated by the constraint of closed cycles. 
The derivative at the end of the {\em adiabat} $\frac{d L}{d C}$ changes sign at this point inducing a change of shape.
The change from concave to convex  cycles in the ${\cal S}_E,\Omega$ plane reflects the evolution of coherence during the cycle.
The distance between ${\cal S}_E$ and ${\cal S}_{vn}$  is a direct consequence of the coherence. The transition point, cycle $g$ has constant coherence.
In the concave cycles theminimum coherence is in the switching points between the {\em adiabats} and the {\em isochores}. At these points
$\frac{ d{\cal C}}{dt} > 0 $ after the {\em isochore } and $\frac{ d{\cal C}}{dt} < 0 $ before the {\em isochore}.
For the convex cycles the maximum coherence is in the middle of the {\em adiabat} with opposite derivatives.

\begin{figure}[tb]
\vspace{0.2cm} 
\center{\includegraphics[height=7.5cm]{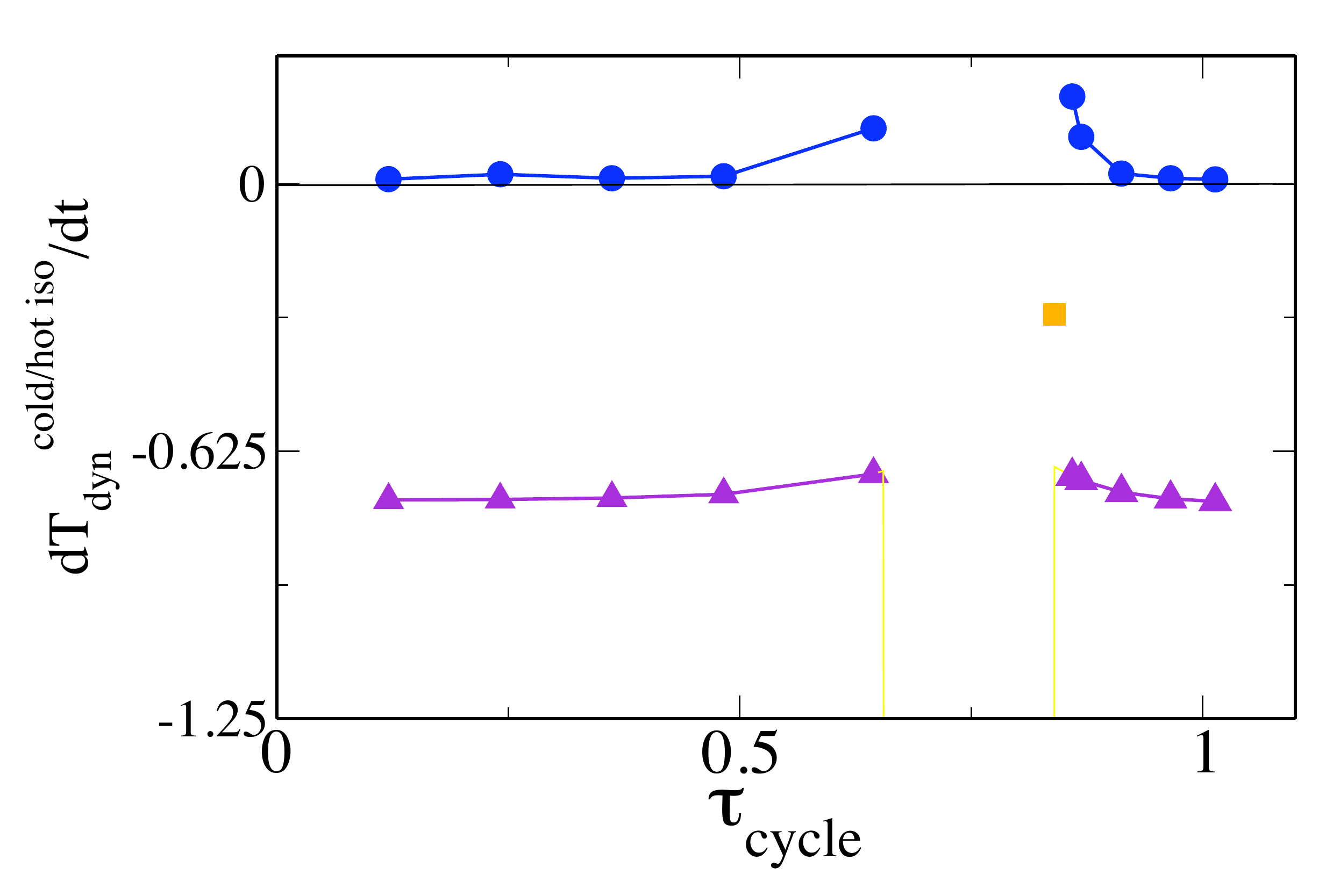}}
\vspace{0.1cm} 
\caption{The time derivative of the dynamical temperature $\frac{d T_{dyn}}{dt} $ on the cold (upper graph)
and hot  (lower graph)  {\em isochore} vs the cycle period $\tau_{cyc}$. The region with missing points corresponds
to the short circuit cycles. The lone point displayed is located in the range of the graph.}
\label{fig:Tdyn}  
\end{figure}

We now turn to the transition points between refrigerators and short circuit cycles.
In a refrigerator the temperature of the working medium should increase in the cold bath and 
$T_{dyn}<T_c$ and decrease in the hot bath $T_{dyn} > T_h$.
Figure \ref{fig:Tdyn} shows the derivative of the dynamical temperature $\frac{d T_{dyn}}{dt} $ 
on the hot and cold {\em isochores} for different cycle times, where the dynamical temperature is defined as \cite{k190}:
\begin{equation}
T_{dyn}~~=~~
\frac {\frac {d { E}}{dt}} {\frac {d { S_E}}{dt}}~~~
\label{eq:dyntemp}
\end{equation} 
The short circuit cycles show no systematic value of $T_{dyn}$ which fluctuates wildly.
This means that the singular change in $T_{dyn}$ is a direct indication of transition from a refrigerator
to short circuit cycle. For a cycle to operate as a refrigerator the working medium should be able to shuttle energy from the cold to the hot bath.
This operation requires a change in internal energy on the {\em adiabat}.
The short circuit cycles are characterized by
small changes in $H$ and therefore they cannot shuttle heat from the cold bath.

One can examine additional special points. For example $l=1$ on both {\em adiabats} corresponds to a frictionless cycle where the connection points $A,B,C,D$ Fig. \ref{fig:1} reside on the energy shell. For $l=3/2$ large coherence is expected and found, resulting in a short circuit cycle.

Finally we address the extremely short cycles which converge when $\tau_{cyc} \rightarrow 0$ to a finite cooling power.
These cycles belong to a universal class of cycles with small action \cite{raam2015}. 
The action for each segment is defined in units of $\hbar$ as $ \int_0^{\tau_j} {\cal L}_j dt $
where ${\cal L}$ is defined in Eq. (\ref{eq:heisenberg}).
For the thermalization {\em isochoric} segments for all cycles studied the action is very small.
On the {\em adiabatic} segments the action is equivalent to the  change in angle $\phi=q \Theta$ which vanishes
when $\mu \rightarrow \infty$. The trivial scaling of energy $\Omega_c/\Omega_h$ does not account since it commutes with all propagators and cancels out.
These cycles are unique in that their cooling power is the result of the coherence. 
Once dephasing is imposed the cooling power vanishes.

\section{ Summary and outlook}
\label{summ2}

Quantum refrigerators are surprisingly diverse. Small changes in parameters 
can lead to a drastic change from  a refrigerator to a short circuit device. In trying to rationalise
this diversity two extreme limits can be identified. Cycle periods approaching infinity lead
to cycles dominated by equilibrium properties. The working medium at all times stays on the energy shell.
The other extreme is infinitely short cycles characterised by vanishing action. These cycles are universal
and possess finite cooling power \cite{raam2015}.

Between these limits the diversity of types and their sensitivity to small changes in parameters
pose a difficult challenge. The present study was devoted to a particular type of sudden Otto refrigerator cycles
where the cycle time is comparable or shorter than the internal time scale.
In these cycles the dynamics is dominated by the {\em adiabatic } segments. These quantum cycles
are characterised by large coherence.  This coherences are accountable for the limit of finite cooling power when 
$\tau_{cyc} \rightarrow 0$. Stochastic cycles conversely have a minimum cycle time \cite{k152}.

For this class of sudden cycles we were able to decipher the  relation between geometry, 
performance and cycle period.
In order for the cycle to shuttle heat the working medium temperature has to be lower than
the cold bath and to heat up by compression to a higher temperature than the hot bath (Cf. Fig. \ref{fig:alltraj}).
We found that this property depends crucially on the geometry of the cycle in the $H,L,C$ vector space.
This topology in turn is determined by the half integer quantiztion condition on the {\em adiabat}.

\subsection*{Aknowledgements}
We want to thank Amikam Levy, and Raam Uzdin  for their help , support and helpful discussions.
This work is supported by the Israel Science Foundation and 
by the European COST network MP1209.

%\subsection*{Bibliography}
%\bibliographystyle{unsrt}

\bibliography{dephc1,pub}

\begin{thebibliography}{27}
\expandafter\ifx\csname natexlab\endcsname\relax\def\natexlab#1{#1}\fi
\expandafter\ifx\csname bibnamefont\endcsname\relax
  \def\bibnamefont#1{#1}\fi
\expandafter\ifx\csname bibfnamefont\endcsname\relax
  \def\bibfnamefont#1{#1}\fi
\expandafter\ifx\csname citenamefont\endcsname\relax
  \def\citenamefont#1{#1}\fi
\expandafter\ifx\csname url\endcsname\relax
  \def\url#1{\texttt{#1}}\fi
\expandafter\ifx\csname urlprefix\endcsname\relax\def\urlprefix{URL }\fi
\providecommand{\bibinfo}[2]{#2}
\providecommand{\eprint}[2][]{\url{#2}}

\bibitem[{\citenamefont{{Ronnie Kosloff}}(2013)}]{k281}
\bibinfo{author}{\bibnamefont{{Ronnie Kosloff}}}, \bibinfo{journal}{Entropy}
  \textbf{\bibinfo{volume}{15}}, \bibinfo{pages}{2100} (\bibinfo{year}{2013}).

\bibitem[{\citenamefont{Allahverdyan et~al.}({2011})\citenamefont{Allahverdyan,
  Hovhannisyan, Janzing, and Mahler}}]{alla11}
\bibinfo{author}{\bibfnamefont{A.~E.} \bibnamefont{Allahverdyan}},
  \bibinfo{author}{\bibfnamefont{K.~V.} \bibnamefont{Hovhannisyan}},
  \bibinfo{author}{\bibfnamefont{D.}~\bibnamefont{Janzing}}, \bibnamefont{and}
  \bibinfo{author}{\bibfnamefont{G.}~\bibnamefont{Mahler}},
  \bibinfo{journal}{Phys. Rev. E} \textbf{\bibinfo{volume}{{84}}},
  \bibinfo{pages}{{041109}} (\bibinfo{year}{{2011}}).

\bibitem[{\citenamefont{{N. Linden, S. Popescu, P.
  Skrzypczyk}}(2010)}]{popescu10}
\bibinfo{author}{\bibnamefont{{N. Linden, S. Popescu, P. Skrzypczyk}}},
  \bibinfo{journal}{Phys.Rev.Lett.} \textbf{\bibinfo{volume}{105}},
  \bibinfo{pages}{130401} (\bibinfo{year}{2010}).

\bibitem[{\citenamefont{Quan et~al.}(2007)\citenamefont{Quan, Liu, Sun, and
  Nori}}]{quan2007}
\bibinfo{author}{\bibfnamefont{H.}~\bibnamefont{Quan}},
  \bibinfo{author}{\bibfnamefont{Y.-x.} \bibnamefont{Liu}},
  \bibinfo{author}{\bibfnamefont{C.}~\bibnamefont{Sun}}, \bibnamefont{and}
  \bibinfo{author}{\bibfnamefont{F.}~\bibnamefont{Nori}},
  \bibinfo{journal}{Physical Review E} \textbf{\bibinfo{volume}{76}},
  \bibinfo{pages}{031105} (\bibinfo{year}{2007}).

\bibitem[{\citenamefont{Brunner et~al.}(2014)\citenamefont{Brunner, Huber,
  Linden, Popescu, Silva, and Skrzypczyk}}]{brunner2014}
\bibinfo{author}{\bibfnamefont{N.}~\bibnamefont{Brunner}},
  \bibinfo{author}{\bibfnamefont{M.}~\bibnamefont{Huber}},
  \bibinfo{author}{\bibfnamefont{N.}~\bibnamefont{Linden}},
  \bibinfo{author}{\bibfnamefont{S.}~\bibnamefont{Popescu}},
  \bibinfo{author}{\bibfnamefont{R.}~\bibnamefont{Silva}}, \bibnamefont{and}
  \bibinfo{author}{\bibfnamefont{P.}~\bibnamefont{Skrzypczyk}},
  \bibinfo{journal}{Physical Review E} \textbf{\bibinfo{volume}{89}},
  \bibinfo{pages}{032115} (\bibinfo{year}{2014}).

\bibitem[{\citenamefont{Correa et~al.}(2014)\citenamefont{Correa, Palao,
  Alonso, and Adesso}}]{correa2014}
\bibinfo{author}{\bibfnamefont{L.~A.} \bibnamefont{Correa}},
  \bibinfo{author}{\bibfnamefont{J.~P.} \bibnamefont{Palao}},
  \bibinfo{author}{\bibfnamefont{D.}~\bibnamefont{Alonso}}, \bibnamefont{and}
  \bibinfo{author}{\bibfnamefont{G.}~\bibnamefont{Adesso}},
  \bibinfo{journal}{Scientific reports} \textbf{\bibinfo{volume}{4}}
  (\bibinfo{year}{2014}).

\bibitem[{\citenamefont{{Tova Feldmann, Eitan Geva, Ronnie Kosloff and Peter
  Salamon}}(1996)}]{k116}
\bibinfo{author}{\bibnamefont{{Tova Feldmann, Eitan Geva, Ronnie Kosloff and
  Peter Salamon}}}, \bibinfo{journal}{{Am. J. Phys.}}
  \textbf{\bibinfo{volume}{64}}, \bibinfo{pages}{485} (\bibinfo{year}{1996}).

\bibitem[{\citenamefont{{Tova Feldmann and Ronnie Kosloff}}(2000)}]{k152}
\bibinfo{author}{\bibnamefont{{Tova Feldmann and Ronnie Kosloff}}},
  \bibinfo{journal}{Phys. Rev. E} \textbf{\bibinfo{volume}{61}},
  \bibinfo{pages}{4774} (\bibinfo{year}{2000}).

\bibitem[{\citenamefont{{Ronnie Kosloff and Tova Feldmann}}(2002)}]{k176}
\bibinfo{author}{\bibnamefont{{Ronnie Kosloff and Tova Feldmann}}},
  \bibinfo{journal}{Phys. Rev. E} \textbf{\bibinfo{volume}{65}},
  \bibinfo{pages}{055102 1} (\bibinfo{year}{2002}).

\bibitem[{\citenamefont{{Tova Feldmann and Ronnie Kosloff}}(2003)}]{k190}
\bibinfo{author}{\bibnamefont{{Tova Feldmann and Ronnie Kosloff}}},
  \bibinfo{journal}{Phys. Rev. E} \textbf{\bibinfo{volume}{68}},
  \bibinfo{pages}{016101} (\bibinfo{year}{2003}).

\bibitem[{\citenamefont{{Yair Rezek, Peter Salamon, Karl Heinz Hoffmann and
  Ronnie Kosloff}}(2009)}]{k243}
\bibinfo{author}{\bibnamefont{{Yair Rezek, Peter Salamon, Karl Heinz Hoffmann
  and Ronnie Kosloff}}}, \bibinfo{journal}{Euro. Phys. Lett.}
  \textbf{\bibinfo{volume}{85}}, \bibinfo{pages}{30008} (\bibinfo{year}{2009}).

\bibitem[{\citenamefont{{Tova Feldmann and Ronnie Kosloff}}(2010)}]{k251}
\bibinfo{author}{\bibnamefont{{Tova Feldmann and Ronnie Kosloff}}},
  \bibinfo{journal}{Euro. Phys. Lett.} \textbf{\bibinfo{volume}{89}},
  \bibinfo{pages}{20004} (\bibinfo{year}{2010}).

\bibitem[{\citenamefont{{Ronnie Kosloff and Tova Feldmann}}(2010)}]{k258}
\bibinfo{author}{\bibnamefont{{Ronnie Kosloff and Tova Feldmann}}},
  \bibinfo{journal}{Phys. Rev. E} \textbf{\bibinfo{volume}{82}},
  \bibinfo{pages}{011134} (\bibinfo{year}{2010}).

\bibitem[{\citenamefont{{G. Thomas and R. S. Johal}}(2011)}]{thomas11}
\bibinfo{author}{\bibnamefont{{G. Thomas and R. S. Johal}}},
  \bibinfo{journal}{Phys. Rev. E} \textbf{\bibinfo{volume}{83}},
  \bibinfo{pages}{031135} (\bibinfo{year}{2011}).

\bibitem[{\citenamefont{Albayrak}(2013)}]{albayrak2013}
\bibinfo{author}{\bibfnamefont{E.}~\bibnamefont{Albayrak}},
  \bibinfo{journal}{International Journal of Modern Physics B}
  \textbf{\bibinfo{volume}{27}} (\bibinfo{year}{2013}).

\bibitem[{\citenamefont{Huang et~al.}(2014)\citenamefont{Huang, Liu, Wang, and
  Niu}}]{huang2014}
\bibinfo{author}{\bibfnamefont{X.}~\bibnamefont{Huang}},
  \bibinfo{author}{\bibfnamefont{Y.}~\bibnamefont{Liu}},
  \bibinfo{author}{\bibfnamefont{Z.}~\bibnamefont{Wang}}, \bibnamefont{and}
  \bibinfo{author}{\bibfnamefont{X.}~\bibnamefont{Niu}}, \bibinfo{journal}{The
  European Physical Journal Plus} \textbf{\bibinfo{volume}{129}},
  \bibinfo{pages}{1} (\bibinfo{year}{2014}).

\bibitem[{\citenamefont{del Campo}(2013)}]{del2013shortcuts}
\bibinfo{author}{\bibfnamefont{A.}~\bibnamefont{del Campo}},
  \bibinfo{journal}{Physical Review Letters} \textbf{\bibinfo{volume}{111}},
  \bibinfo{pages}{100502} (\bibinfo{year}{2013}).

\bibitem[{\citenamefont{Jarzynski}(2013)}]{jarzynski2013generating}
\bibinfo{author}{\bibfnamefont{C.}~\bibnamefont{Jarzynski}},
  \bibinfo{journal}{Physical Review A} \textbf{\bibinfo{volume}{88}},
  \bibinfo{pages}{040101} (\bibinfo{year}{2013}).

\bibitem[{\citenamefont{Schmiedl and Seifert}(2008)}]{schmiedl2008}
\bibinfo{author}{\bibfnamefont{T.}~\bibnamefont{Schmiedl}} \bibnamefont{and}
  \bibinfo{author}{\bibfnamefont{U.}~\bibnamefont{Seifert}},
  \bibinfo{journal}{EPL (Europhysics Letters)} \textbf{\bibinfo{volume}{81}},
  \bibinfo{pages}{20003} (\bibinfo{year}{2008}).

\bibitem[{\citenamefont{{Tova Feldmann and Ronnie Kosloff}}(2004)}]{k201}
\bibinfo{author}{\bibnamefont{{Tova Feldmann and Ronnie Kosloff}}},
  \bibinfo{journal}{Phys. Rev. E} \textbf{\bibinfo{volume}{70}},
  \bibinfo{pages}{046110} (\bibinfo{year}{2004}).

\bibitem[{\citenamefont{Feldmann and Kosloff}(2012)}]{k274}
\bibinfo{author}{\bibfnamefont{T.}~\bibnamefont{Feldmann}} \bibnamefont{and}
  \bibinfo{author}{\bibfnamefont{R.}~\bibnamefont{Kosloff}},
  \bibinfo{journal}{Phys. Rev. E} \textbf{\bibinfo{volume}{85}},
  \bibinfo{pages}{051114} (\bibinfo{year}{2012}),
  \urlprefix\url{http://link.aps.org/doi/10.1103/PhysRevE.85.051114}.

\bibitem[{\citenamefont{Alicki and Lendi}(1987)}]{alicki87}
\bibinfo{author}{\bibfnamefont{R.}~\bibnamefont{Alicki}} \bibnamefont{and}
  \bibinfo{author}{\bibfnamefont{K.}~\bibnamefont{Lendi}},
  \emph{\bibinfo{title}{{Quantum Dynamical Semigroups and Applications }}}
  (\bibinfo{publisher}{Springer-Verlag, Berlin}, \bibinfo{year}{1987}).

\bibitem[{\citenamefont{Lindblad}(1976)}]{lindblad76}
\bibinfo{author}{\bibfnamefont{G.}~\bibnamefont{Lindblad}},
  \bibinfo{journal}{Comm. Math. Phys.} \textbf{\bibinfo{volume}{48}},
  \bibinfo{pages}{119} (\bibinfo{year}{1976}).

\bibitem[{\citenamefont{{Uri Banin, Allon Bartana, Sanford Ruhman and Ronnie
  Kosloff}}(1994)}]{k108}
\bibinfo{author}{\bibnamefont{{Uri Banin, Allon Bartana, Sanford Ruhman and
  Ronnie Kosloff}}}, \bibinfo{journal}{J. Chem. Phys.}
  \textbf{\bibinfo{volume}{101}}, \bibinfo{pages}{8461} (\bibinfo{year}{1994}).

\bibitem[{\citenamefont{Girolami}(2014)}]{girolami2014observable}
\bibinfo{author}{\bibfnamefont{D.}~\bibnamefont{Girolami}},
  \bibinfo{journal}{Physical review letters} \textbf{\bibinfo{volume}{113}},
  \bibinfo{pages}{170401} (\bibinfo{year}{2014}).

\bibitem[{\citenamefont{Plastina et~al.}(2014)\citenamefont{Plastina, Alecce,
  Apollaro, Falcone, Francica, Galve, Gullo, and Zambrini}}]{plastina2014}
\bibinfo{author}{\bibfnamefont{F.}~\bibnamefont{Plastina}},
  \bibinfo{author}{\bibfnamefont{A.}~\bibnamefont{Alecce}},
  \bibinfo{author}{\bibfnamefont{T.}~\bibnamefont{Apollaro}},
  \bibinfo{author}{\bibfnamefont{G.}~\bibnamefont{Falcone}},
  \bibinfo{author}{\bibfnamefont{G.}~\bibnamefont{Francica}},
  \bibinfo{author}{\bibfnamefont{F.}~\bibnamefont{Galve}},
  \bibinfo{author}{\bibfnamefont{N.~L.} \bibnamefont{Gullo}}, \bibnamefont{and}
  \bibinfo{author}{\bibfnamefont{R.}~\bibnamefont{Zambrini}},
  \bibinfo{journal}{Physical Review Letters} \textbf{\bibinfo{volume}{113}},
  \bibinfo{pages}{260601} (\bibinfo{year}{2014}).

\bibitem[{\citenamefont{Uzdin et~al.}(2015)\citenamefont{Uzdin, Levy, and
  Kosloff}}]{raam2015}
\bibinfo{author}{\bibfnamefont{R.}~\bibnamefont{Uzdin}},
  \bibinfo{author}{\bibfnamefont{A.}~\bibnamefont{Levy}}, \bibnamefont{and}
  \bibinfo{author}{\bibfnamefont{R.}~\bibnamefont{Kosloff}},
  \bibinfo{journal}{arXiv preprint arXiv:1502.06592}  (\bibinfo{year}{2015}).

\end{thebibliography}

\end{document}